\documentstyle[twocolumn,prc,aps,epsfig]{revtex}
\input epsf

\begin{document}

\twocolumn[\hsize\textwidth\columnwidth\hsize\csname @twocolumnfalse\endcsname

\title {
   {Flow at the SPS and RHIC as a Quark-Gluon Plasma Signature} 
}
\author {D. Teaney$^1$, J. Lauret$^2$, E.V.~Shuryak$^1$}
\address {
$^1$ Department of Physics and Astronomy, $^2$ Department of Chemistry, \\
State University of New York at Stony Brook, NY 11794-3800, 
}
\date{\today}
\maketitle    
\begin{abstract}
Radial and elliptic flow in non-central heavy ion collisions 
can constrain the effective Equation of State(EoS) of the excited nuclear matter.
To this end, a model combining relativistic hydrodynamics and
a hadronic transport code(RQMD\cite{Sorge-RQMD}) is developed. 
For an EoS with a first order phase transition, the model
reproduces both the radial and elliptic flow data at the SPS. 
With the EoS fixed from SPS data, we quantify predictions at RHIC where the
Quark Gluon Plasma(QGP) pressure is expected to drive additional 
radial and elliptic flow.  Currently, the strong elliptic flow 
observed in the first RHIC measurements does not conclusively signal 
this nascent QGP pressure. 
Additional measurements are suggested to pin down the EoS.
\end{abstract}

\vspace{0.1in}
]

\begin{narrowtext}   
\newpage
1. By colliding heavy nuclei at the SPS and RHIC accelerating facilities,
physicists hope to excite hadronic matter into  a new
phase consisting of deconfined 
quarks and gluons -- the  Quark Gluon Plasma(QGP)\cite{QGP-Reviews}.
After the collision, the produced particles move collectively 
or $flow$ and this flow may  quantify the effective Equation of State(EoS) of the matter.
In central PbPb collisions at the SPS, a strong radial flow 
is observed\cite{RadialFlow-Review}. The matter develops a collective 
transverse velocity approaching (1/2)c.
In non-central collisions, a radial and an $elliptic$ flow are
observed\cite{EllipticFlow-Review,NA49-BDependence,STAR-Elliptic}.
Since in non-central collisions 
the initial nucleus-nucleus overlap region has an elliptic
shape, the initial pressure gradient 
is larger along the impact parameter and the matter moves preferentially
in this direction\cite{Ollitrault-Elliptic}.
%

The phase transition to the QGP  influences both the radial and
elliptic flows. QCD lattice simulations
show an approximately 1st order phase
transition\cite{Lattice-Eos}. Over a wide range of energy 
densities $e=.5-1.4\,GeV/fm^3$, 
the temperature and pressure are nearly constant. Over this range then,
the ratio of pressure to energy density $p/e$, decreases and  
reaches a minimum at a particular energy density
 known as the {\em softest point}, $e_{sp}\approx1.4\,GeV/fm^{3}$\cite{SoftestPoint}.
When the initial energy density is close to $e_{sp}$,   
the small pressure (relative to $e$) cannot
effectively  accelerate the matter.
However, when the initial energy density is well above $e_{sp}$, 
$p/e$ approaches 1/3, and the larger pressure 
drives collective 
motion
\cite{SoftestPoint,Kataja-MixedPhase}.
At a time of $\sim1\,fm/c$, the energy densities at the SPS($\sqrt
s_{NN}=17\,GeV$) and RHIC ($\sqrt s_{NN}=130\,GeV$) 
are very approximately
4 and $7\,GeV/fm^3$ 
respectively\cite{NA49-EnergyDensity,Phobos-Multiplicity}. 
Based on these experimental estimates, 
the hard QGP phase is expected to live significantly 
longer at RHIC than at the SPS.
The
final flows of the produced particles should reflect this difference.
In this paper we pose the question: Can 
both the  radial and elliptic flow at the SPS and RHIC be described
by a single effective EoS?
%

Since the various hadron species have different elastic cross sections,
they freezeout (or decouple) from the hot fireball at different times\cite{Sorge-Strange}.
Because flow builds up over time, it is essential to model this differential
freezeout. It was ignored in previous
hydrodynamic simulations of non-central heavy ion collisions and elliptic
flow was over-predicted 
flow by a factor of two\cite{Kolb-Flow,Kolb-UU}.

2. The Hydro to Hadrons(H2H) 
model  will be described in detail elsewhere\cite{H2H-Flow}. Other
authors have previously constructed a similar model for central
collisions\cite{HydroUrqmd}. 
The model evolves the QGP and mixed phases as a
relativistic fluid, 
but switches to a hadronic cascade (RQMDv2.4\cite{Sorge-RQMD})
at the beginning of the hadronic phase
to model differential freezeout.
The computer code consists of three distinct components. 
Assuming Bjorken scaling, the first component solves the equations of 
relativistic hydrodynamics in the transverse plane\cite{Ollitrault-Elliptic} and
constructs a switching surface at a temperature, $T_{switch}=160\,MeV$. 
The second component generates hadron on the switching surface 
using the Cooper-Frye formula\cite{CooperFrye} with 
 a theta function rejecting backward going particles
\cite{Freezeout-Theta,Teaney-RTTC}.  Finally, the third
component (RQMD) sequentially re-scatters the generated hadrons until
freezeout. 
\begin{figure}[h]
\begin{center}
   \vspace{-0.2in}
   \includegraphics[width=3.3in,height=3.0in]{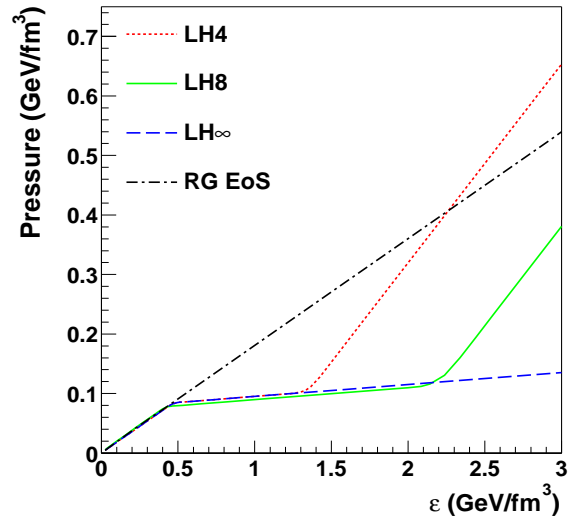}
\end{center}
   \caption[]{
      \label{psCs2}   
      The pressure versus the energy density($\epsilon$) for
      different EoSs (see text).
      EoSs with Latent Heats $0.4\,GeV/fm^{3}$, $0.8\,GeV/fm^{3}$,...
are labeled as LH4, LH8,...etc. 
   }
\end{figure}

For the hydrodynamic evolution, a family of EoSs 
was constructed with an adjustable Latent Heat(LH)
\linebreak[4]
\begin{figure}[h]
\begin{center}
   \vspace{-.4in}
   \includegraphics[width=3.3in,height=3.0in]{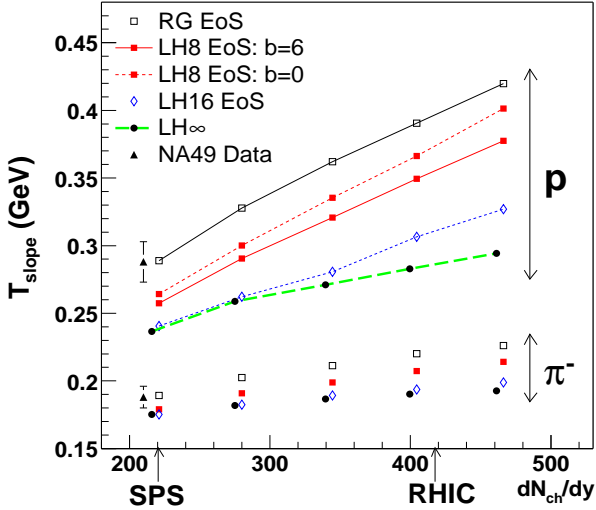}
\end{center}
   \caption[]{
   \label{psTemperaturedNdy}
The transverse mass slope($T_{slope}$) as a function
of the total charged particle multiplicity in PbPb collisions 
at an impact parameter of b=6\,fm\,(see also \cite{Kataja-MixedPhase,HydroUrqmd}). 
For consistency with the elliptic study in Fig.~\ref{psPionV2dNdy},
we show b=6\,fm 
although the NA49 data points\cite{NA49-Slopes}
are for the 5\% most central events, or b$<$3.5 fm.  
For all EoSs at the SPS, the proton slope parameters 
at b=6\,fm are $\approx7\,MeV$ 
smaller than at b=0\,fm, as for the b=0 LH8 curve. 
The difference is negligible for $\pi^{-}$. 
   }
\end{figure}
\noindent
(see Fig.~\ref{psCs2}).  LH$\infty$ is considered as a limiting
case, mimicking non-equilibrium phenomena\cite{Sorge-PreEquilibrium}.
The hadron phase exists up to a critical temperature of $T_{c}=165\,MeV$, 
and consists of ideal gas mixture of the
 meson pseudoscalar and vector nonets and the baryon octet
and decuplet.
The hadron phase is followed
by a mixed phase with a specified LH, which is finally followed by a QGP phase with 
$C_s^2 = 1/3$.  
In addition, a Resonance Gas(RG) EoS
was constructed with a constant speed of sound above the hadron phase.

3.  {\em Radial flow} is quantified experimentally 
by slope parameters, $T_{slope}$;
the momentum spectrum of each particle is fit  to  the form 
$dN/dM_T^{2}\,dy|_{y=0}=C\,e^{-M_T/T_{slope}}$ where $M_T^2=P_T^2+m^2$.
$T_{slope}$ incorporates random thermal motion
and the collective transverse velocity. 

In Fig.~\ref{psTemperaturedNdy}, the pion and proton slope parameters 
are plotted as functions of the  
total charged particle multiplicity in the collision.
Look first at the leftmost points at SPS multiplicities and 
compare the model and experimental slopes:
The proton slope data favor a relatively hard EoS -- LH8 or harder. 
A direct comparison
of the model to published spectra \cite{NA49-Spectra} 
supports this claim\cite{H2H-Flow}. 
A RG EoS can also reproduce the proton flow.
A similar analysis of elliptic flow (shown and quantified below)
favors a relatively soft EoS -- LH8 or softer. With some caveats, LH8 
represents a happy middle which can reproduce
both the radial and  elliptic flow at the SPS. 

Look now at the energy/multiplicity dependence of the slopes.  
For all EoSs, $T_{slope}$ increases with the 
collision energy\cite{Kataja-MixedPhase,HydroUrqmd}. 
For a  soft EoS (e.g. LH$\infty$) the
increase is small  and for a hard EoS (e.g. LH8) the increase
is large. At RHIC multiplicities, the difference between
the slope parameters is large and easily experimentally observable.

4. {\em Elliptic flow} is quantified experimentally by 
the elliptic flow parameter, 
$v_{2}=\langle \cos(2\Phi) \rangle$; 
 here $\Phi$ is the 
angle around the beam measured relative to the  impact parameter
and  $\langle\,\rangle$ denotes an average over the
single particle distribution, $\frac{dN}{dP_{T}d\Phi}$. 
$v_{2}(P_{T})$  is found by holding $P_{T}$  
constant in while averaging $\cos(2\Phi)$ over $\frac{dN}{dP_{T}d\Phi}$.
$v_{2}$ measures the response of the fireball to the
the spatial deformation of the overlap region,  
which is usually quantified in a Glauber model\cite{SN402}
by the eccentricity 
$\epsilon = \langle\langle y^{2} - x^{2}\rangle\rangle/\langle\langle x^{2} + y^{2}\rangle\rangle$.  
Since the response($v_2$) is proportional to
the driving force($\epsilon$), the ratio   
$v_{2}/\epsilon$ is used to compare different impact 
parameters and nuclei\cite{Heiselberg-99,Voloshin-LowDensity}.

In Fig.~\ref{psPionV2dNdy}(a), the number elliptic flow ($v_{2}$)
is plotted as a function of charged particle multiplicity at an 
impact parameter of 6\,fm. 
Before studying the energy dependence, look at the 
magnitude of the elliptic flow at the SPS.
For LH8, the stars 
show the pion $v_{2}$ when the matter is evolved as a fluid 
until a decoupling temperature of $T_{f}=120\,MeV$; 
they illustrate
the excessive elliptic flow typical of pure hydrodynamics.
Once a cascade
is included, LH8 (the squares) is only $\approx20\%$ above the data --
a substantial improvement.
Typically in
hydrodynamic calculations, the freezeout temperature $T_{f}$ is
adjusted to fit the proton $P_{T}$ spectrum. However, 
protons are driven by a pion ``wind'' and  decouple  from the fireball  
$5\,fm/c$ after the pions on average.  This pion wind 
accounts for the strong
proton flow at the SPS and is not described by ideal 
hydrodynamics\cite{HydroUrqmd,H2H-Flow}. 
In order to match the observed proton flow, hydrodynamic calculations
must decouple at low freezeout temperatures,
$T_{frz} \approx120\,MeV/c$. This low temperature has  
two consequences for elliptic flow: First
the reduction of 
elliptic flow due to resonance
decays is small $\approx15\%$, compared to $\approx30\%$ in the H2H model.
Second, compared to a cascade, 
the hydrodynamics generates twice as much elliptic
flow during the late cool hadronic stages of the evolution.
By including
the pion ``wind'', and more generally by decoupling differentially,
we can simultaneously describe the radial and  
elliptic flow data at the SPS.

The energy dependence of $v_{2}$ 
is the central issue.
As seen in Fig.~\ref{psPionV2dNdy}, the H2H model predicts an  
increase in elliptic flow by a factor    
$\approx1.4$ and is in reasonable agreement with SPS and RHIC  
flow data. This result was presented prior to the publication of
RHIC data \cite{Teaney-RTTC}.
In contrast, UrQMD, a hadronic cascade based on string dynamics, 
predicts a decrease by a factor of  $\approx2$ \cite{UrQMD-Elliptic}. 
This is because the UrQMD string model has 
a super-soft EoS at
high energies\cite{UrQMD-Hagedorn}. For pure hydrodynamics 
as illustrated by
the stars, 
$v_{2}$ is approximately constant\cite{Kolb-Flow} (but see \cite{Kolb-UU}).
For HIJING\cite{HIJING}, a model which considers 
only the initial parton collisions, 
$v_{2}$ is $\approx0$\cite{HIJING-Elliptic}. 
The first RHIC data clearly contradict these  models. 

The increase in $v_2$ is now used to constrain the EoS of the excited 
matter.
The QCD phase diagram has two 
\begin{figure}[t]
\begin{center}
   \includegraphics[width=3.3in,height=3.0in]{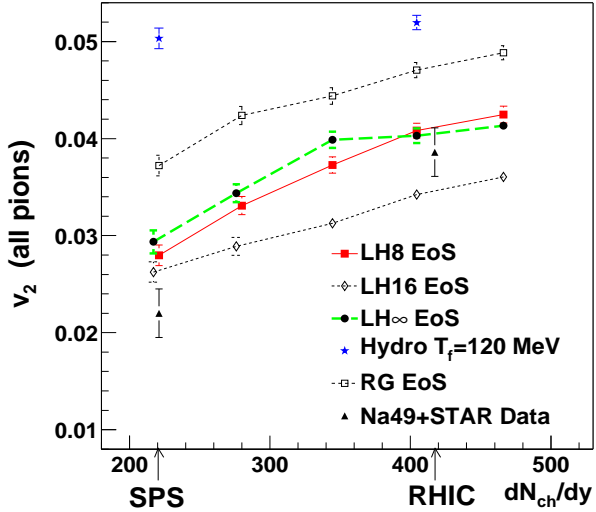}
\end{center}
   \caption[]{
   \label{psPionV2dNdy}
(a)\,The number elliptic flow   
parameter $v_2$ 
as a function of the charged particle multiplicity in
 PbPb collisions at a an impact parameter of b=6\,fm. 
At the SPS, the NA49 $v_2$  
data point is extrapolated to b=6\,fm using
Fig.~3 in \cite{NA49-BDependence}.  
At RHIC, the STAR $v_2$ data point is extrapolated to 
$N_{ch}/N_{ch}^{max}=0.545$ (b=6\,fm in AuAu) using  Fig.~3 in 
\cite{STAR-Elliptic}.  
The comparison to data is a little unfair: 
For the model, $v_{2}$ is calculated using   
all pions in PbPb collisions. 
For the NA49 data, $v_{2}$  is measured using 
only $\pi^{-}$ in PbPb (a -3\% correction to the model). 
For the STAR data, $v_{2}$ is measured using 
charged hadrons in AuAu (a +5\% correction to the model).
   }
\end{figure}
\noindent 
distinguishing features. 
It is soft at low energy densities and subsequently hard at high energies. 
A RG EoS(the open squares) 
has no softness and the elliptic flow is clearly too strong 
both at the SPS and RHIC. 
The entire family of EoSs, LH8 through LH$\infty$,  
reproduces the elliptic flow data
in both energy regimes. Counter-intuitively, as the latent heat
is increased, $v_{2}$ first decreases and then increases. In the
final count,
LH8 and LH$\infty$ have roughly the same $v_{2}$. 
However, they develop the  $v_{2}$ in  
different ways.
For LH8, the EoS shifts from hard to soft and the
early pressure starts an early elliptic expansion.
For LH$\infty$, the EoS is just soft and the elliptic
expansion stalls. However because the expansion 
is stalled, the LH$\infty$ collision lifetime($\approx13\,fm/c$ at RHIC) 
is significantly longer than the LH8 lifetime($\approx9\,fm/c$ at
RHIC)\cite{SoftestPoint}. 
Over the long LH$\infty$ lifetime,  $v_{2}^{LH\infty}$ 
steadily grows and is finally comparable to $v_{2}^{LH8}$.
As the latent heat is increased from LH8 to LH16, 
the EoS becomes becomes softer and $v_{2}$ at first decreases.
However, as the EoS is made   
softer still, the lifetime increases and $v_{2}$ rises again.

5.{\em Impact Parameter Dependence}.
In Fig.~\ref{psBscanV2particip}, 
$v_{2}$ for LH8 as a function of the number participants($N_{p}$) 
is compared to data. 
Different EoSs show a similar participant (or b) dependence.
The agreement is good at RHIC where the multiplicity 
is high. 
For ideal hydrodynamics, $v_{2} \propto \epsilon \propto (N_{p}^{max}-N_{p})$\cite{Ollitrault-Elliptic}.
In the low density limit, since the
\linebreak[4]
\begin{figure}[t]
\begin{center}
   \includegraphics[width=3.3in, height=3.0in]{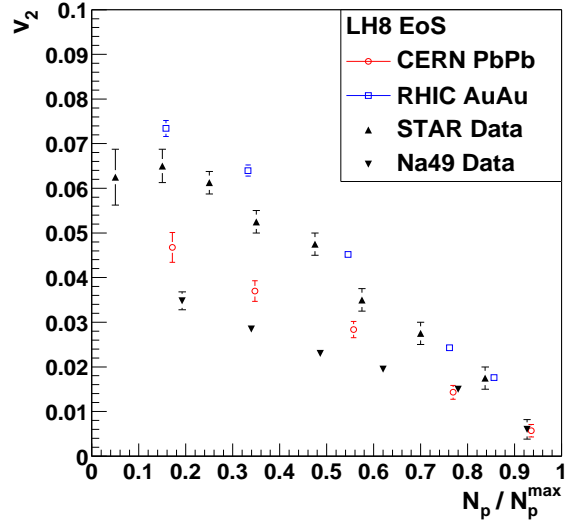}
\end{center}
   \caption[]{
      \label{psBscanV2particip}
$v_2$  
versus the number of participants($N_{p}$) relative to the maximum .
The model and the NA49 $v_{2}$ values\cite{NA49-BDependence} 
at the SPS are for $\pi^{-}$. 
The NA49 data are mapped from b to participants using\cite{NA49-Glauber}.
The model and the STAR $v_{2}$ values\cite{STAR-Elliptic} at RHIC are
for charged particles. The model does not include weak decays. 
The number of charged particles is assumed proportional to $N_{p}$.
   }
\end{figure}
\noindent
response is proportional to
the number of collisions, $v_{2} \propto \epsilon \frac{dN}{dy} 
\propto (N_{p}^{max} -N_{p})N_{p}$.
Therefore, $v_{2}$ has a different 
$N_{p}$(or b) dependence in the hydrodynamic and low density 
limits\cite{Heiselberg-99,Voloshin-LowDensity}.
At RHIC, except in very peripheral collisions,  
the $N_{p}$ dependence   
is clearly linear  
and strongly supports the hydrodynamic limit\cite{STAR-Elliptic}. 
At the SPS, the $N_{p}$ dependence may not be clearly linear,  
but it also does not  follow the low density limit. 
Two-pion correlations, 
may change the data analysis\cite{Ollitrault-CorrelationsA}, 
 reduce $v_{2}$ in the periphery and improve the low density agreement. 

Finally in Fig.~\ref{psV2DataB}, 
$v_{2}$ is studied both as a function
of transverse momentum and impact parameter.  
For both LH8 and LH$\infty$, the
calculation produces too much elliptic flow in peripheral 
collisions(45-85\%), and too $little$ elliptic flow in the
most central collisions(0-11\%). 
The $P_{T}$ dependence of $v_{2}$ also clarifies the
difference between and LH8 and LH$\infty$: LH$\infty$,
a super soft EoS, generates elliptic flow only at low momentum
while LH8, a hard EoS, generates elliptic at high momentum.


6.{\em Summary and Discussion}. 
By incorporating differential freezeout, the Hydro to Hadrons(H2H) 
model simultaneously reproduces the radial and elliptic flow at 
the SPS and  RHIC. 
At the SPS, the radial flow demands 
an EoS with a latent heat $LH \gtrsim 0.8\,GeV/fm^{3}$, 
while elliptic flow demands
an EoS with a latent  heat $LH \lesssim 0.8\,GeV/fm^{3}$. 
Further, in contrast to string and collision-less parton models,
the  increase in $v_{2}$ 
is naturally explained using hydrodynamics. This
challenges the prevailing 
view\cite{STAR-Elliptic,Voloshin-LowDensity} 
that the SPS is in the low density
regime and that the increase in $v_{2}$ represents a transition 
\linebreak[4]
\begin{figure}[t]
\begin{center}
   \vspace{-.4in}
   \includegraphics[width=3.3in,height=3.0in]{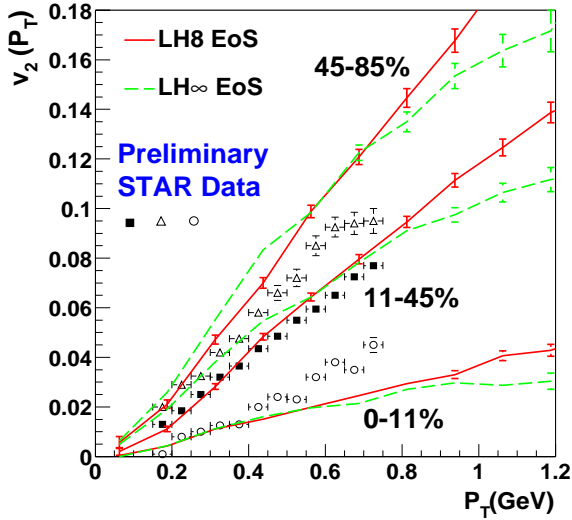}
\end{center}
  \caption[]{
   \label{psV2DataB}
Elliptic flow of charged pions as  a function of $P_{T}$
and centrality for AuAu collisions at RHIC.
For centrality selections  
The percentages shown 0-11\%, 11-45\% and 45-85\%, indicate
the fraction of the total geometric cross section for three
centrality selections 
 0\,fm$<$b$<$4.2\,fm,
 4.2\,fm$<$b$<$8.4\,fm and
 8.4\,fm$<$b$<$11.6\,fm.
The preliminary data points were presented in\cite{QM2001,Snellings-QM2001}.  
The model curves were found by parameterizing
the model data points and averaging over the specified impact
range with the geometric weight, $2\pi b\,db$. 
   } 
\end{figure}
\noindent
to the hydrodynamic regime. However, 
the  increase in $v_{2}$ does not uniquely signal the 
asymptotic QGP pressure. 
Indeed, at RHIC collision energies, 
a very soft EoS can have the same $v_{2}$ as an EoS with
a well developed QGP phase. This EoS is not academic
since softness can mimic 
non-equilibrium phenomena\cite{Sorge-PreEquilibrium}.
To reveal the underlying EoS and the burgeoning QGP pressure,
the collision energy should be scanned from the SPS to RHIC.
If the prevailing low density view of the SPS is correct, a
transition in the b dependence of elliptic flow should be observed 
over the energy range\cite{Heiselberg-99,Voloshin-LowDensity}. 
In addition, for different EoSs,
$v_{2}$ depends differently 
on collision energy and transverse momentum 
(Fig.~\ref{psPionV2dNdy} and Fig.~\ref{psV2DataB}).
Taken with the radial flow\,(Fig.~\ref{psTemperaturedNdy}), 
this experimental 
information would help settle the EoS of hot hadronic matter.

{\bf Acknowledgments}.
The work is partly supported by
the US DOE grant No. DE-FG02-88ER40388
and grant No. DE-FGO2-87ER 40331. 

{\bf Note Added}.  After the submission of this work,
the STAR and PHENIX collaborations reported proton at anti-proton
spectra\cite{QM2001}. The preliminary 
spectra favor LH8-LH16 and disfavor LH$\infty$\cite{Teaney-QM2001}.
%
%
%
%
%

\end{narrowtext}
\end{document}